\documentclass{article}
\usepackage[a4paper, total={6in, 8in}]{geometry}

\usepackage{graphicx}
\usepackage{subfigure}
\usepackage{colortbl}
\usepackage{amsmath}
\usepackage{amsfonts} 
\usepackage{tabularx}

\usepackage{array}
\usepackage{booktabs}

\usepackage{tikz}
\usepackage{quantikz}
\usepackage{authblk}

\usepackage[colorlinks,citecolor=blue,linkcolor=red,urlcolor=blue]{hyperref}

\newcommand{\V}{$\boldsymbol{\vee}$}
\newcommand{\X}{$\cdot$}

\usepackage{algorithm}
\usepackage{algpseudocode}

\title{TQml Simulator: \\optimized simulation of quantum machine learning}

\author{Viacheslav Kuzmin}
\author{Basil Kyriacou}
\author{Tatjana Protasevich}
\author{Mateusz Papierz}
\author{Mo~Kordzanganeh}
\author{Alexey~Melnikov}

\affil{Terra Quantum AG, Kornhausstrasse 25, 9000 St.~Gallen, Switzerland}

\begin{document}

\maketitle

\begin{abstract}
Hardware-efficient circuits employed in Quantum Machine Learning are typically composed of alternating layers of uniformly applied gates. High-speed numerical simulators for such circuits are crucial for advancing research in this field. In this work, we numerically benchmark universal and gate-specific techniques for simulating the action of layers of gates on quantum state vectors, aiming to accelerate the overall simulation of Quantum Machine Learning algorithms. Our analysis shows that the optimal simulation method for a given layer of gates depends on the number of qubits involved, and that a tailored combination of techniques can yield substantial performance gains in the forward and backward passes for a given circuit. Building on these insights, we developed a numerical simulator, named TQml Simulator, that employs the most efficient simulation method for each layer in a given circuit. We evaluated TQml Simulator on circuits constructed from standard gate sets, such as rotations and CNOTs, as well as on native gates from IonQ and IBM quantum processing units. In most cases, our simulator outperforms equivalent Pennylane's default.qubit simulator by up to a factor of 10, depending on the circuit, the number of qubits, the batch size of the input data, and the hardware used.

\end{abstract}

\section{Introduction}

Quantum Machine Learning (QML)~\cite{schuld2021machine, melnikov2023quantum, acampora2025quantum} is a dynamic and rapidly expanding field at the intersection of quantum computing and machine learning, attracting much attention from researchers. While in standard ML exploiting classical hardware such as CPU, GPU, or TPU, data are transformed primarily through explicit matrix–matrix multiplications, QML encodes the data into a quantum state and processes it via (in most scenarios) the unitary evolution of a quantum circuit~\cite{benedetti2019parameterized}. Recent theoretical and numerical studies~\cite{liu2021rigorous, gyurik2023exponential} have demonstrated the possibility of obtaining quantum advantage with this approach on synthetic datasets; however, achieving such advantages in real-world applications remains a significant challenge that requires further exploration~\cite{cerezo2023does, gil2024relation}. 

Given the current limitations of quantum hardware~\cite{kordzanganeh2023benchmarking}, much of QML research relies on numerical simulations of quantum devices. Among the various simulation methods, state-vector simulation~\cite{gangapuram2024benchmarking,guerreschi2020intel} has emerged as the most widely used approach due to its superior performance for small system sizes and its conceptual similarity to classical machine learning workflows, where feature vectors are propagated through successive layers of trainable operations. Furthermore, this method readily facilitates automatic gradient computation through backpropagation when integrated with modern machine learning frameworks such as PyTorch~\cite{Paszke2017AutomaticDI}, Jax~\cite{jax2018github}, or TensorFlow~\cite{abadi2016tensorflowlargescalemachinelearning}. Since training a QML model involves multiple iterations of forward and backward passes, the simulation time of a quantum ansatz circuit is a critical factor. In this work, we investigate approaches for optimized state-vector simulation within the QML framework.

The state vector of an \( n \)-qubit quantum system is a vector \(\psi\) of \(2^n\) complex numbers, with each element representing the probability amplitude of a corresponding basis state in the \(2^n\)-dimensional Hilbert space. The evolution of the quantum state can be straightforwardly described by the multiplication of a \(2^n \times 2^n\) unitary matrix \(U\) with the state vector, \(U\psi\), an operation that incurs a computational complexity of \(\mathcal{O}(2^{2n})\), see Fig.~\ref{fig:U_and_Einsum}(a). In this work, we explore alternative simulation approaches that leverage additional information specific to each individual quantum gate, as given in Table~\ref{tbl:gates}. This extra knowledge includes: 
\begin{itemize}
    \item Layer-wise gates appearance in QML circuits,
    \item The locality of the gates,
    \item The diagonality of the corresponding matrices,
    \item The fact that some transformations are represented by real-valued matrices,
    \item The effective representation of a gate’s action as a permutation.
\end{itemize}
Exploiting these properties allows reducing the complexity of gate application to \(\mathcal{O}(n2^n)\) or even \(\mathcal{O}(2^n)\). For instance, one can leverage gate locality by applying local gates using the Einstein summation (Einsum) technique, as illustrated in Fig.~\ref{fig:U_and_Einsum}(b). A notable example of using such operation-specific knowledge is a development of a QAOA-focused simulator~\cite{lykov2023fast} that precomputes the diagonal of the problem Hamiltonian to accelerate simulation, ultimately solving certain problems faster than state-of-the-art classical solvers~\cite{shaydulin2024evidence}. This method is also exploited by a popular PennyLane default.qubit simulator~\cite{bergholm2022pennylaneautomaticdifferentiationhybrid}.

Here, we aim to develop an optimized quantum simulator specifically for QML applications~\cite{kurkin2025forecasting,sagingalieva2025hybrid,sagingalieva2025photovoltaic}. For each gate layer, we benchmarked a set of applicable approaches and found that the optimal simulation method for a particular layer often depends on the number of qubits in the circuit. Based on this observation, we propose an approach to build an "Optimized" simulator of a QML circuit that exploits the optimal simulation techniques chosen for each gate layer individually in order to minimize the forward or combined forward and backward execution times for the available hardware, such as multithreaded CPUs, GPUs, or TPUs. Our approach plays a role similar to that of modern software compilers that target a specific hardware architecture, but it acts one layer higher: instead of rearranging low-level instructions as, e.g, hardware-centric quantum-simulation compilers do~\cite{guerreschi2022fast}, it chooses the most efficient algorithmic recipe for each gate layer for given computational resources. We named the obtained simulator as Terra Quantum Machine Learning (TQml) Simulator. By implementing TQml Simulator using a PyTorch back-end, we show that it outperforms PennyLane default.qubit simulator by up to a factor of 10, depending on the circuit, the number of qubits, the batch size of the input data, and the hardware used.
The remainder of the paper is structured as follows. In Sec.~\ref{techniques}, we present the simulation techniques for various layers of gates and provide numerical benchmark results comparing these techniques. In Sec.~\ref{optimal_sim}, we describe our method for creating the optimized TQml Simulator with PyTorch back-end for QML models and benchmark its performance against the PennyLane default.qubit simulator using a representative circuit, evaluating both input data parallelization and full training iteration (forward and backward passes). Then, in Sec.~\ref{sec:tqml-jax} we also benchmark our TQml simulator with JAX back-end and analyze its performance in comparison with PyTorch back-end and other simulators. Finally, in Sec.~\ref{conclusion}, we summarize our findings and conclude our work.
    
\section{Techniques for Applying Layers of Gates to a State Vector}
\label{techniques}

\subsection{Methods}
In this section, we discuss various approaches for simulating gate layers and benchmark them by comparing them against each other and the PennyLane default.qubit simulator. In all simulations, including those run with the default.qubit simulator, we employ a PyTorch back-end and set the precision to \texttt{complex128} (since using \texttt{complex64} in PennyLane is non-trivial). 

We chose the default.qubit simulator over the more advanced lightning.qubit simulator because, like our code, it is entirely written in Python. Although lightning.qubit exploits C++ to achieve faster simple forward passes, its execution time scales linearly when batching input data, as if the data were processed sequentially, eventually underperforming compared to the default.qubit simulator. Furthermore, lightning.qubit does not support backpropagation for gradient computations and relies on the adjoint method~\cite{jones2020efficientcalculationgradientsclassical}, which, in our tests, underperformed compared to backpropagation. These factors motivated our decision to use the default.qubit simulator for a fair comparison.

For benchmarking, we use a warmup technique, where each method is run several times before measurements are taken. Reported times are averaged over 10 iterations, with the standard deviation indicated by shaded areas in the plots. Note that data for different plots might have been collected on different machines, but all data within an individual plot were obtained using the same machine.

In the work, we consider the gates given in Table~\ref{tbl:gates} with their properties discussed further in the text. The matrix representations of the considered gates are provided in Appendix~\ref{Matrix Representations}

\begin{table}[h!]
\centering

\definecolor{lightgray}{gray}{0.9}

\begin{tabular}{lccccc}
\toprule
\textbf{Gate} & \textbf{(Anti)diagonal} & \textbf{Permutation} & \textbf{Real} & \textbf{Parametrized} & \textbf{N qubits} \\
\midrule
Z     & \V & \X & \V & \X & 1 \\
\rowcolor{lightgray} CZ    & \V & \X & \V & \X & 2 \\
Rz    & \V & \X & \X & \V & 1 \\
\rowcolor{lightgray} Rzz   & \V & \X & \X & \V & 2 \\
GPI   & \V & \X & \X & \X & 1 \\
\rowcolor{lightgray} S     & \V & \X & \X & \X & 1 \\
CNOT  & \X & \V & \V & \X & 2 \\
\rowcolor{lightgray} X     & \X & \V & \V & \X & 1 \\
H     & \X & \X & \V & \X & 1 \\
\rowcolor{lightgray} Ry    & \X & \X & \V & \V & 1 \\
Rx    & \X & \X & \X & \V & 1 \\
\rowcolor{lightgray} Rot   & \X & \X & \X & \V & 1 \\
GPI2  & \X & \X & \X & \V & 1 \\
\rowcolor{lightgray} MS    & \X & \X & \X & \V & 2 \\
Sx    & \X & \X & \X & \X & 1 \\
\rowcolor{lightgray} ECR   & \X & \X & \X & \X & 2 \\
\bottomrule
\end{tabular}
\caption{Quantum gates and their properties. M\o lmer--S\o rensen (MS) gate, GPI, and GPI2 are native gates of IonQ Aria quantum computer. ECR is a native gate of IBM Eagle-type quantum processors.  }
\label{tbl:gates}
\end{table}

\subsection{Universal Approaches}

    \begin{figure}[]
        \centering
        \includegraphics[width=0.8\linewidth]{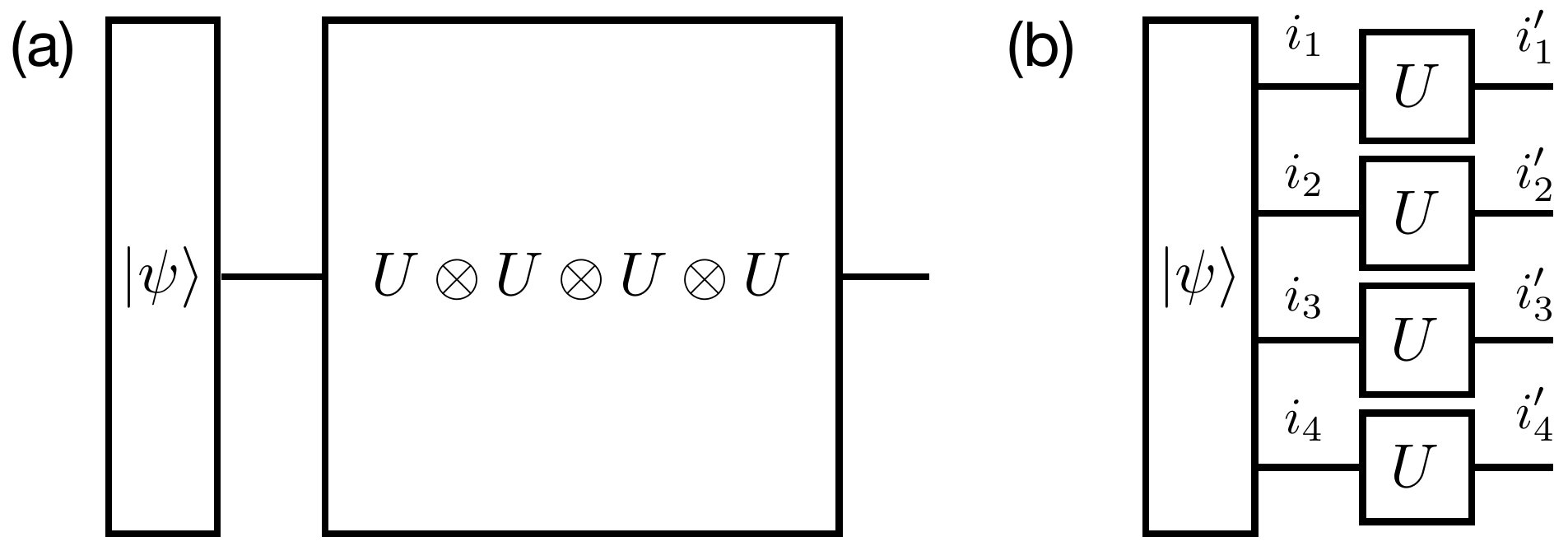}
        \caption{Tensor diagrams illustrating two approaches for applying a layer of single-qubit gates, \((\prod_i U^{(i)})|\psi\rangle\) (where the superscript denotes the qubit index). (a) The explicit unitary operation is represented as \((\bigotimes_{i=1}^4 U^{(i)}) \times \psi\), while (b) the Einstein summation approach is depicted by \(U^{i'_1}_{i_1} U^{i'_2}_{i_2} U^{i'_3}_{i_3} U^{i'_4}_{i_4} \, \psi_{i_1 i_2 i_3 i_4}\). In these diagrams, boxes denote tensors and wires represent their indices; connected wires indicate summation over the corresponding indices.}
        \label{fig:U_and_Einsum}
    \end{figure}

    \begin{figure}[]
        \centering
        \includegraphics[width=5.9in]{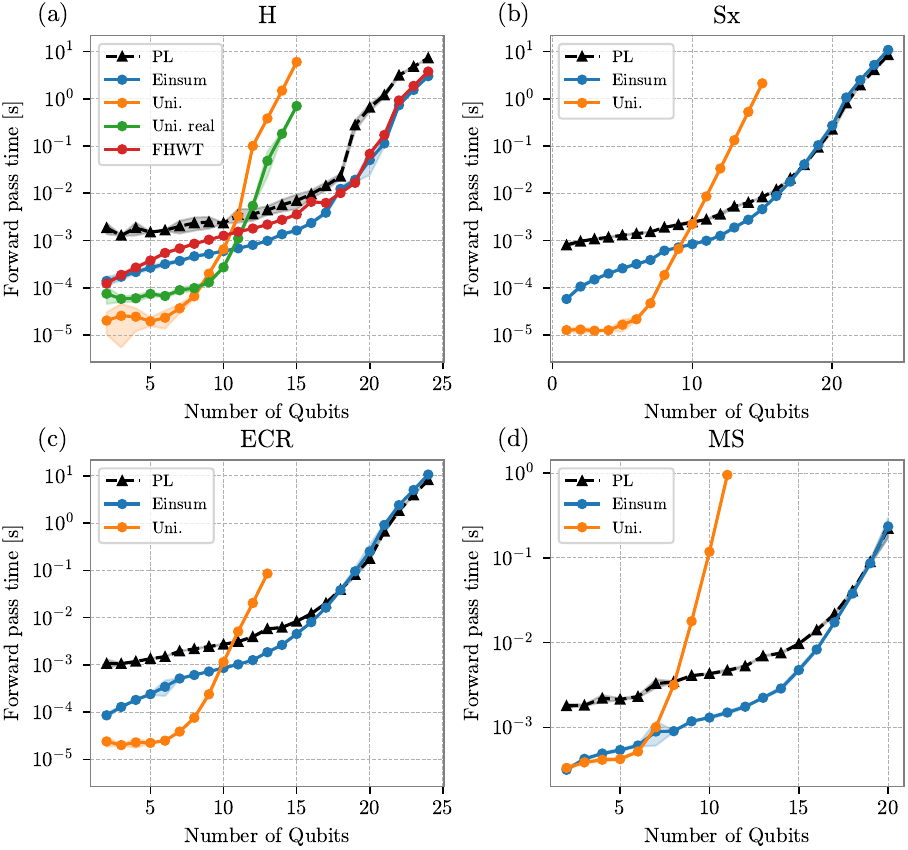}
\caption{Forward pass time on a single CPU thread for layers composed of unparametrized (a) H, (b) Sx, and (c) native IBM ECR gates and (d) parametrized MS gate as a function of the number of qubits. Results are obtained using the PennyLane default.qubit simulator (PL) and the methods evaluated in this work: Einsum, Unitary operation (Uni.), Unitary real operation (Uni. real), and Fast Hadamard-Walsh Transform (FHWT).}
        \label{fig:H_Sx_Ect}
    \end{figure}

In this section, we explore several simulation techniques for applying a layer of quantum gates to a quantum state in a manner that is independent of the specific gate type.

\subsubsection{Unitary Operation}

Consider an \( n \)-qubit quantum system whose state is represented by a vector \( \psi \) with \( 2^n \) complex elements, each corresponding to the probability amplitude of a basis state in a \( 2^n \)-dimensional Hilbert space. The evolution of this state under any unitary transformation $U$ is described by
\[
\psi' = U \times \psi,
\]
where \( U \) is a \( 2^n \times 2^n \) unitary matrix, see Fig.~\ref{fig:U_and_Einsum}(a). This direct matrix-vector multiplication, with a computational complexity of \(\mathcal{O}(2^{2n})\), is the most straightforward method, also from a numerical point of view. Despite its poor scaling relative to other techniques discussed in this work, its simplicity makes it the fastest approach for systems with up to 7–9 qubits, as demonstrated in Figs.~\ref{fig:H_Sx_Ect}-\ref{fig:X_CNOT}.

For cases where \( U \) is a real-valued matrix, the operation can be efficiently implemented by decomposing the state vector into its real and imaginary components:
\[
\psi' = U \times \mathrm{Re}(\psi) + i\,U \times \mathrm{Im}(\psi).
\]
This approach effectively halves the computational cost compared to the complex-valued operation. Figure~\ref{fig:H_Sx_Ect} demonstrates that this approach has superior performance for implementing the layer of H gates on 9 and 10 qubits.

\subsubsection{Sequential Einsum}

Hardware-efficient circuit ansätze commonly used in QML applications~\cite{leone2024practical} rely on local quantum operations, such as single-qubit rotations and two-qubit entangling gates, that are applied uniformly across all qubits. In this context, applying such local gates using Einstein summation can be preferred over other approaches.

To do so numerically, the state vector is reshaped into a tensor \(\psi_{i_1 i_2 \dots i_n}\), where each index \(i_k\) (with dimension 2) corresponds to an individual qubit, as illustrated in Fig.~\ref{fig:U_and_Einsum}(b). A local unitary operation acting on \(l\) qubits can similarly be expressed as tensor with \(2l\) indices, each of dimension $2$. For example, a 2-qubit gate represented by \( U^{i'_k i'_m}_{i_k i_m} \) can be expressed as a tensor with 4 indices and shape $[2,2,2,2]$. Applying such an operation to the state tensor involves performing Einstein summation over the indices associated with the qubits being operated on:
\[
\psi'_{i_1 \dots i'_k \dots i'_m \dots i_n} = U^{i'_k i'_m}_{i_k i_m} \, \psi_{i_1 i_2 \dots i_n}.
\]
The computational complexity for applying a single \(l\)-local operation in this way is \(\mathcal{O}(2^{l+n})\).

If we assume that the number of operations in a single layer scales linearly with the number of qubits \(n\), the overall complexity for applying an entire layer via the Einsum approach becomes \(\mathcal{O}(n \cdot 2^{l+n})\). For the typical cases of 1- and 2-qubit operations (\(l=1\) or \(l=2\)), this results in \(\mathcal{O}(n2^n)\) complexity. This is significantly lower than the \(\mathcal{O}(2^{2n})\) complexity incurred by applying the full unitary operation, yielding a substantial efficiency gain. Benchmarking our custom-optimized Einsum technique, we observed that, in most cases (as illustrated in Figs.~\ref{fig:H_Sx_Ect}–\ref{fig:RY_RX_Rot}), the performance matches that of the PennyLane library in the limit of large qubit numbers, while outperforming it for smaller qubit counts, likely due to specific implementation optimizations.
    
\subsection{Permutation Gates}

    \begin{figure}[]
        \centering
        \includegraphics[width=5.9in]{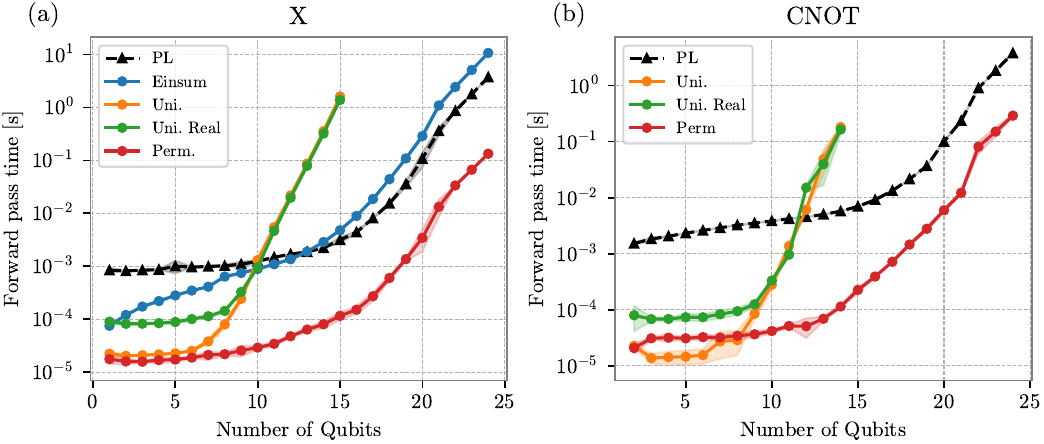}
\caption{Forward pass time on a single CPU thread for layers composed of unparametrized permutation gates (a) X and (b) CNOT as a function of the number of qubits. Results are obtained using the PennyLane default.qubit simulator (PL) and the methods evaluated in this work: Einsum, Unitary operation (Uni.), Unitary real operation (Uni. real), and Permutation (Perm.).}
        \label{fig:X_CNOT}
    \end{figure}

Permutation gates in quantum computing—such as the X, CNOT, and SWAP gates—rearrange the elements of a quantum state vector without altering their values. This unique property allows these operations to be implemented very efficiently in numerical simulations, as they only require reassigning memory pointers instead of performing arithmetic operations.

For example, consider the CNOT gate, which is defined by the matrix
\[
\mathrm{CNOT} = \begin{pmatrix}
1 & 0 & 0 & 0 \\
0 & 1 & 0 & 0 \\
0 & 0 & 0 & 1 \\
0 & 0 & 1 & 0 \\
\end{pmatrix}.
\]
In a two-qubit system, this gate permutes the third and fourth elements of the state vector. This permutation can be described by a permutation vector, \(\sigma_{\mathrm{CNOT}}\), where each entry indicates the new position of the corresponding probability amplitude. For the CNOT gate, the permutation vector is:
\[
\sigma_{\mathrm{CNOT}} = \begin{pmatrix}
1 & 2 & 4 & 3
\end{pmatrix}.
\]
Thus, the action of the CNOT gate on a state vector \(\phi\) is succinctly written as:
\[
\phi' = \phi[\sigma_{\mathrm{CNOT}}].
\]

Permutation matrices, like diagonal matrices, form a mathematical group under composition. This means that multiple permutation operations can be combined into a single permutation. For instance, a series of CNOT gates arranged in a ring to entangle all qubits can be represented as a single ``permutation layer''. 

Because these operations involve only the reassignment of indices, their computational complexity is \(\mathcal{O}(2^n)\), making them exceptionally efficient for numerical simulations. This efficiency is confirmed by numerical benchmarks of a circular layer of CNOT gates and a layer of X gates implemented with the permutation techniques, as shown in Fig.~\ref{fig:X_CNOT}.

\subsection{Diagonal gates}

    \begin{figure}[]
        \centering
        \includegraphics[width=5.9in]{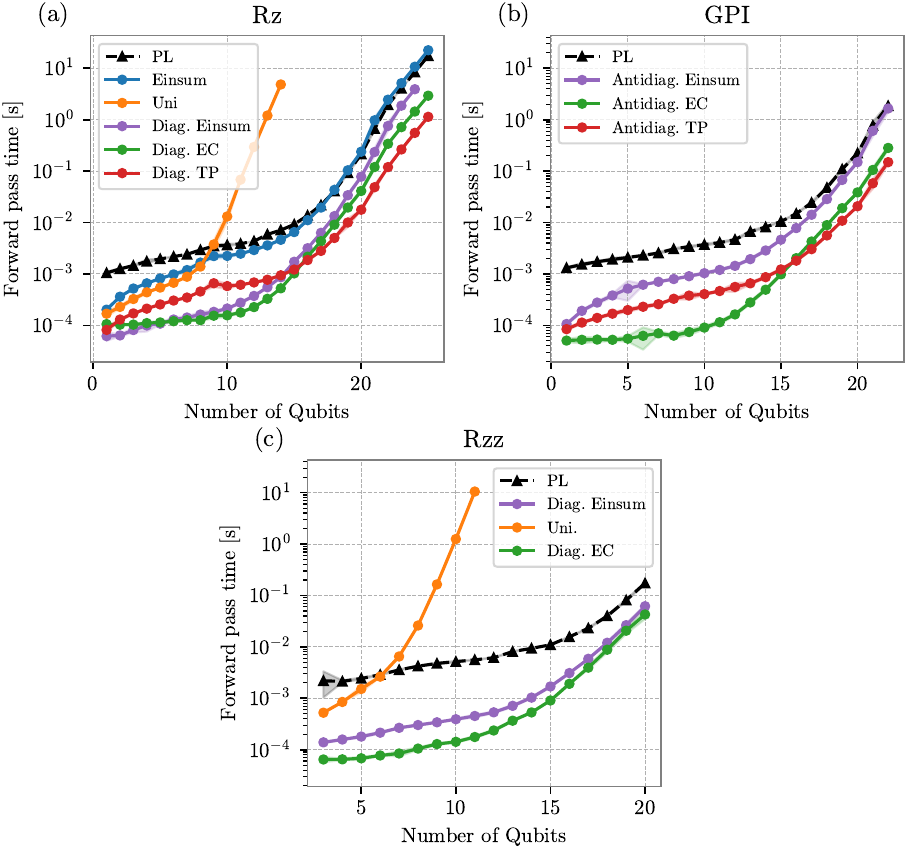}
\caption{Forward pass time on a single CPU thread for layers composed of parametrized diagonal (a) Rz, (c) Rzz, and (b) antidiagonal native IonQ GPI gates versus the number of qubits. Results are obtained using the PennyLane default.qubit simulator (PL) and the techniques evaluated in this work: Einsum, Unitary operation (Uni.), Diagonal Einsum (Diag. Einsum), Eigenphase Computation (Diag. EC), and Diagonal Tensor Product (Diag. TP).}
        \label{fig:Rz_GPI_ZZ}
    \end{figure}

In this section, we describe methods for applying a layer of diagonal gates, such as Rz and Rzz, to a quantum state. The techniques discussed can be readily extended to layers of antidiagonal gates—such as the GPI gate, which is native to IonQ Aria processors.

\subsubsection{Eigenphase Computation}

The diagonal matrices affect the state vector through elementwise multiplication by their diagonal. If the diagonal elements are precomputed, transforming the state vector of an $n$-qubit system requires only \(\mathcal{O}(2^n)\) operations. For static gates like \(Z\) or \(S\), the diagonal elements can be stored in memory, allowing repeated applications at \(\mathcal{O}(2^n)\) cost. 

For parameterized diagonal gates like \(Rz(\theta)\), the diagonal elements must be computed on the fly. The \(Rz(\theta)\) gate is unitary, with eigenvalues of the form \(e^{i\alpha}\). The eigenphase angles \(\alpha\) are determined by the parameter \(\theta\). As an example, for an \(Rz\) layer, these angles can be computed through a matrix operation:
\[
\alpha = K_{Rz} \theta,
\]
where \(K_{Rz}\) is a \(2^n \times n\) matrix with entries of \(\pm 1\). Each row of \(K_{Rz}\) represents a unique binary configuration of the \(n\) qubits, with \(+1\) corresponding to a \(0\) and \(-1\) corresponding to a \(1\). To generate \(K_{Rz}\), one starts with a \(2^n \times n\) binary counting matrix \(J\) and applies the elementwise transformation:
\[
K_{Rz} = -2J + 1.
\]
For example, when \( n = 3 \):
\[
J =
\begin{pmatrix}
0 & 0 & 0 \\
0 & 0 & 1 \\
0 & 1 & 0 \\
0 & 1 & 1 \\
1 & 0 & 0 \\
1 & 0 & 1 \\
1 & 1 & 0 \\
1 & 1 & 1 \\
\end{pmatrix}
\quad \longrightarrow \quad
K_{Rz} =
\begin{pmatrix}
+1 & +1 & +1 \\
+1 & +1 & -1 \\
+1 & -1 & +1 \\
+1 & -1 & -1 \\
-1 & +1 & +1 \\
-1 & +1 & -1 \\
-1 & -1 & +1 \\
-1 & -1 & -1 \\
\end{pmatrix}.
\]

For a fixed number of qubits \(n\), \(K_{Rz}\) remains constant and needs to be computed only once. With \(K_{Rz}\) in hand, the action of the \(RZ(\theta)\) layer on a state vector \(\psi\) is given by:
\[
\psi^\prime = \psi \circ \exp(i K_{Rz} \theta),
\]
where \(\circ\) denotes the elementwise product and the exponential is applied elementwise to the vector \(K_{Rz}\theta\).

The overall complexity of forming a diagonal layer using this approach involves three main steps:

\begin{itemize}
\item A matrix-vector multiplication with complexity \(\mathcal{O}(n2^n)\)
\item \(\mathcal{O}(2^n)\) exponential evaluations
\item\(\mathcal{O}(2^n)\) elementwise multiplications.
\end{itemize}
    
Although the overall complexity \(\mathcal{O}(n2^n)\) is the same as the Einsum approach, numerical benchmarks for layers of diagonal gates given in Fig.~\ref{fig:Rz_GPI_ZZ} demonstrate a performance improvement of roughly one order of magnitude.

For a ring of \( R_{zz} \) gates applied to nearest-neighbor pairs—including the pair connecting the last and first qubits—the parity of the bits in each computational basis state must be computed to construct the matrix \( K_{Rzz} \). Specifically, the elements of \( K_{Rzz} \) are defined as
\[
K_{Rzz}[i,j] = K_{Rz}[i,j]\, K_{Rz}\Big[i,(j+1) \bmod n\Big],
\]
where \( K_{Rz} \) is the matrix defined previously. With \( K_{Rzz} \) determined, the action of the \( R_{zz} \) layer on a state vector is performed analogously to the \( R_z \) layer.

To apply a layer of anti-diagonal GPI gates, the same matrix as for Rz is used (i.e., \( K_\mathrm{GPI} \equiv K_{Rz} \)); however, an additional reshuffling of the state vector is required. This is achieved by mapping
\[
\psi^\prime[2^n-1-i] = \Bigl(\psi \circ \exp\bigl(i\, K_\mathrm{GPI}\, \theta\bigr)\Bigr)[i].
\]

\subsubsection{Diagonal Tensor Product}

An alternative approach to computing the diagonal of a multi-qubit diagonal operation is by taking the tensor product of the diagonals of the individual single-qubit gates:
\[
\mathrm{diag}(U) = \bigotimes_i \mathrm{diag}(U^{(i)}),
\]
where the superscript indicates the corresponding qubit. Computing this tensor product requires \(2^{n+1} - 4\) multiplications, which is \(\mathcal{O}(2^n)\). Once the full diagonal is computed, applying the gate layer to the state vector reduces to an elementwise multiplication:
\[
\psi^\prime = \mathrm{diag}(U) \circ \psi,
\]
which also has a complexity of \(\mathcal{O}(2^n)\). Consequently, the overall complexity of this approach is \(\mathcal{O}(2^n)\), making it more efficient than the Eigenphase Computation method. Our benchmarks in Fig.~\ref{fig:Rz_GPI_ZZ} demonstrate that the Diagonal Tensor Product outperforms the Eigenphase Computation technique for a larger number of qubits, confirming the gain in complexity.

\subsubsection{Diagonal Einsum}

A third method extends the Einstein summation approach— used for full unitary operations—to the case of diagonal gates. In this variant, the summation is performed exclusively over the diagonal elements of the local operations. Formally, consider a local operation on two qubits $i$ and $k$ represented by a unitary with diagonal \( D_{i_k\,i_m} \); its application to a state is given by:
\[
\psi'_{i_1\cdots i'_k\cdots i'_m\cdots i_n}
=\;\delta_{i'_k,i_k}\,\delta_{i'_m,i_m}\,D_{i_k\,i_m}\;\psi_{i_1\cdots i_k\cdots i_m\cdots i_n},
\]
with $\delta_{i'_m,i_m}$ the Kronecker delta. Each diagonal Einsum operation has a computational complexity of \(\mathcal{O}(2^n)\). Consequently, when applying a full layer of gates—which typically involves \(\mathcal{O}(n)\) such local operations—the overall complexity becomes \(\mathcal{O}(n2^n)\). This is comparable to the complexity of the Eigenphase Computation method. As shown in Fig.~\ref{fig:Rz_GPI_ZZ}, the diagonal Einsum technique generally achieves similar performance, although it tends to slightly underperform the Eigenphase Computation approach in most cases.

\subsection{H-Rz Expansion}

    \begin{figure}[]
        \centering
            \includegraphics[width=5.9in]{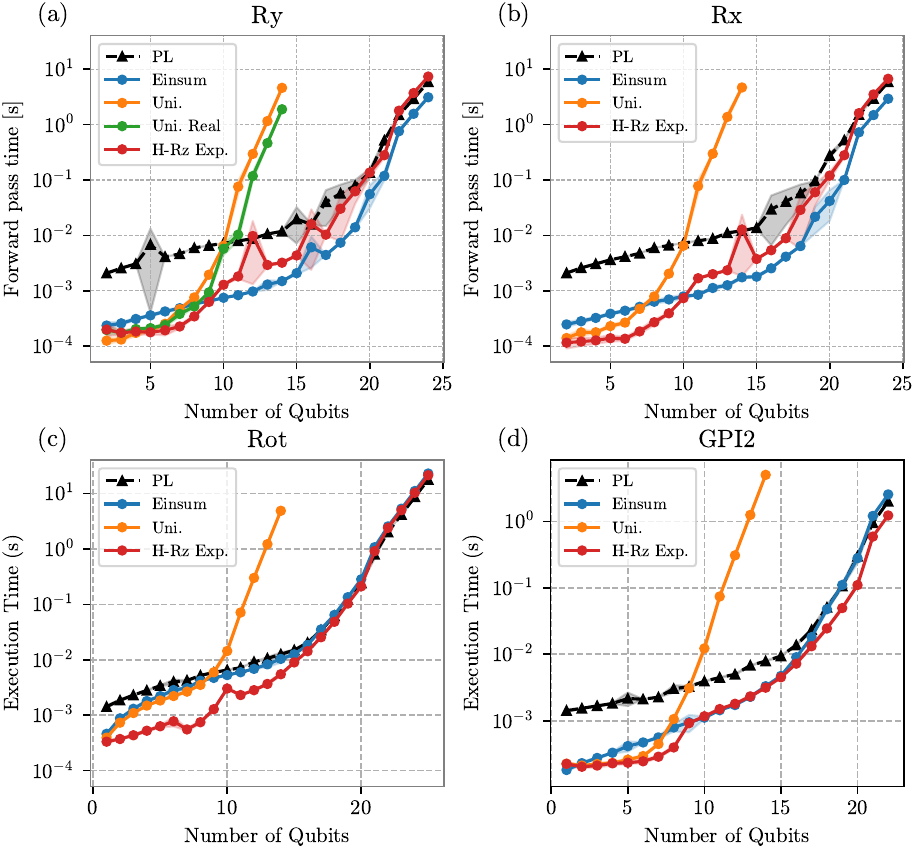}
\caption{Forward pass time on a single CPU thread for layers composed of parametrized rotation gates (a) Ry, (b) Rx, (c) general Rot, and (d) native IonQ GPI2 gates as a function of the number of qubits. Results are obtained using the PennyLane default.qubit simulator (PL) and the methods evaluated in this work: Einsum, Unitary operation (Uni.), Unitary real operation (Uni. real), and H-Rz Expansion (H-Rz Exp.).}
        \label{fig:RY_RX_Rot}
    \end{figure}

For single-qubit rotation gates, we also explored a heuristic approach that expands a rotation into a sequence of Rz and H gates. In particular, we used the following decompositions for \(R_x\), \(R_y\), a general rotation \(R(\phi,\theta,\omega)\), and native IonQ GPI2 gate:
\begin{align}
R_x(\theta) &= H \, R_z(\theta) \, H,\\[1mm]
R_y(\theta) &= Z \, H \, R_z(\theta) \, H \, Z,\\[1mm]
R(\phi, \theta, \omega) &= R_z(\omega+\pi) \, H \, R_z(\theta) \, H \, R_z(\phi+\pi),\\[1mm]
\mathrm{GPI2}(\theta) &= R_z(-\theta) \, R_x(\pi/2) \, R_z(\theta).
\end{align}

In simulating these decompositions, we employed the optimal techniques for a given number of qubits as indicated in Fig.~\ref{fig:H_Sx_Ect} for H layers and in Fig.~\ref{fig:Rz_GPI_ZZ} for Rz layers. Similar to $H$ gate, $R_x(\pi/2)$ in the $\mathrm{GPI2}(\theta)$ decomposition is constant for a given number of qubits, and we used the same optimal techniques except the Real Unitary Operation since $R_x(\pi/2)$ is complex. As shown in Fig.~\ref{fig:RY_RX_Rot}, this H-Rz Expansion method surprisingly outperforms other approaches, such as Einsum, for up to 10 qubits in the case of \(R_y\) and \(R_x\) gates, and it happens to be optimal for a layer of general rotation Rot and GPI2 gates.

In this context, we also investigated the use of the Fast Hadamard-Walsh Transform (FHWT) algorithm for simulating layers of H gates, which has a complexity of \(\mathcal{O}(n 2^n)\). Although FHWT showed performance comparable to the Einsum approach (as illustrated in Fig.~\ref{fig:H_Sx_Ect}), it significantly underperformed when computing gradients via backpropagation (results not shown). 
    
\section{Optimized TQml Simulator for QML applications}
\label{optimal_sim}

\begin{figure}[]
    \centering
    \includegraphics[width=0.6\linewidth]{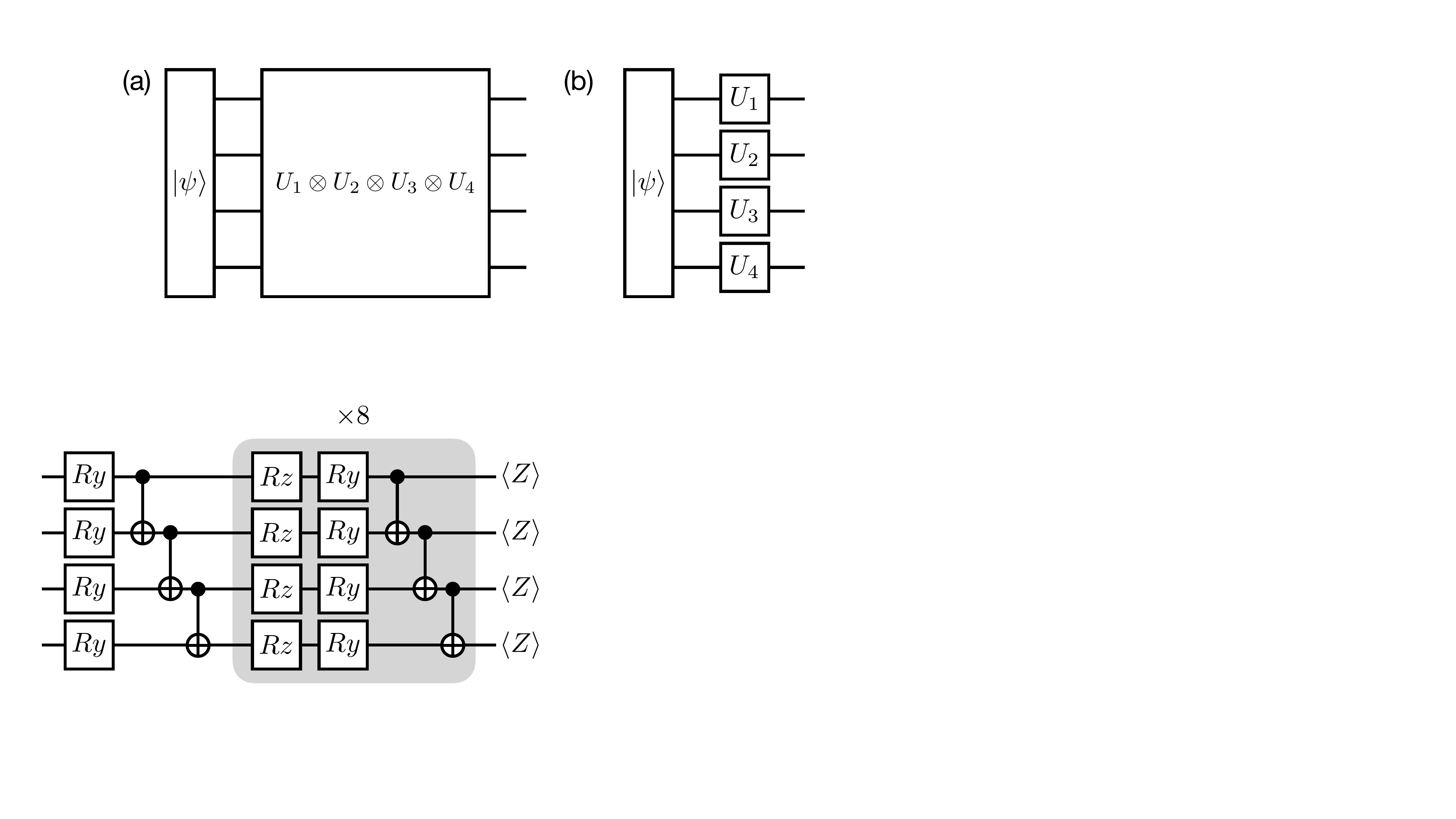}
\caption{4-qubit example of a circuit used for benchmarks reported in Figs.~\ref{fig:batching} and \ref{fig:forward_and_backprop_scaling}. The circuit begins with an initial layer of \(R_y\) rotations and a layer of CNOT gates, followed by a block of \(R_z\)-\(R_y\)-CNOT layers that is repeated 8 times. In Fig.~\ref{fig:batching}, no measurements are performed; in Fig.~\ref{fig:forward_and_backprop_scaling}, single-qubit observables \(\langle Z \rangle\) are measured and summed to yield the output and its derivatives.}
    \label{fig:Circuit_Rz_Ry_Cnot}
\end{figure}

In the previous section, we presented numerical benchmarks for various techniques used to apply different gate layers on a quantum state vector. Our results indicate that for each specific gate, there exists an optimal simulation method that depends on the number of qubits in the circuit. This insight motivates the development of an optimized simulator for QML circuits, which we call the Terra Quantum Machine Learning (TQml) Simulator, that selects the most efficient simulation technique for each layer given specific simulation hardware resources—similar to how a compiler optimizes code for available hardware. The criteria for choosing the optimal technique can include:

\begin{enumerate}
    \item The number of available CPU threads,
    \item The type of available hardware (e.g., CPU, GPU, or TPU),
    \item Memory allocation,
    \item Times for forward and/or backward passes.
\end{enumerate}

Notably, the last criterion implies that the optimal set of simulation techniques may differ between training and inference phases.

For simplicity, in this work we define optimality as the minimal forward pass time on a single CPU thread, as given in Figs.~\ref{fig:H_Sx_Ect}--\ref{fig:RY_RX_Rot}. In the following, we compare the performance of the TQml Simulator, compiled with the given criteria, with the PennyLane default.qubit simulator. As a demonstrative example, we use the circuit shown in Fig.~\ref{fig:Circuit_Rz_Ry_Cnot} built with a typical set of gates \{Ry,Rz,CNOT\}, varying the number of qubits while keeping the number of layers fixed. We also considered hardware-specific circuits designed for IonQ ion-trap and IBM superconducting quantum computers. We provide these extra results in Appendix~\ref{Hardware-specific simulators}.

The first benchmark we consider is the forward pass time as a function of the input batch size. Batching is critical in QML because it allows both software and hardware to parallelize computations over multiple inputs, thereby greatly reducing training and inference times. For this benchmark, we investigated two types of data encoding in the circuit from Fig.~\ref{fig:Circuit_Rz_Ry_Cnot}: 

\begin{itemize}
    \item In the \emph{Vanilla Quantum (VQ)} approach, input data is encoded only in the first \(R_z\) layer. This results in a batched state vector that is then evolved under the same operations for all inputs, analogous to classical ML.
    \item In the \emph{Quantum Depth Infused (QDI)} approach~\cite{sagingalieva2023hybrid}, the data is encoded in every \(R_z\) layer, meaning that the batched state vector undergoes batched encoding at each block, which may affect parallelization behavior.
\end{itemize}

\begin{figure}[]
    \centering
    \includegraphics[width=5.9in]{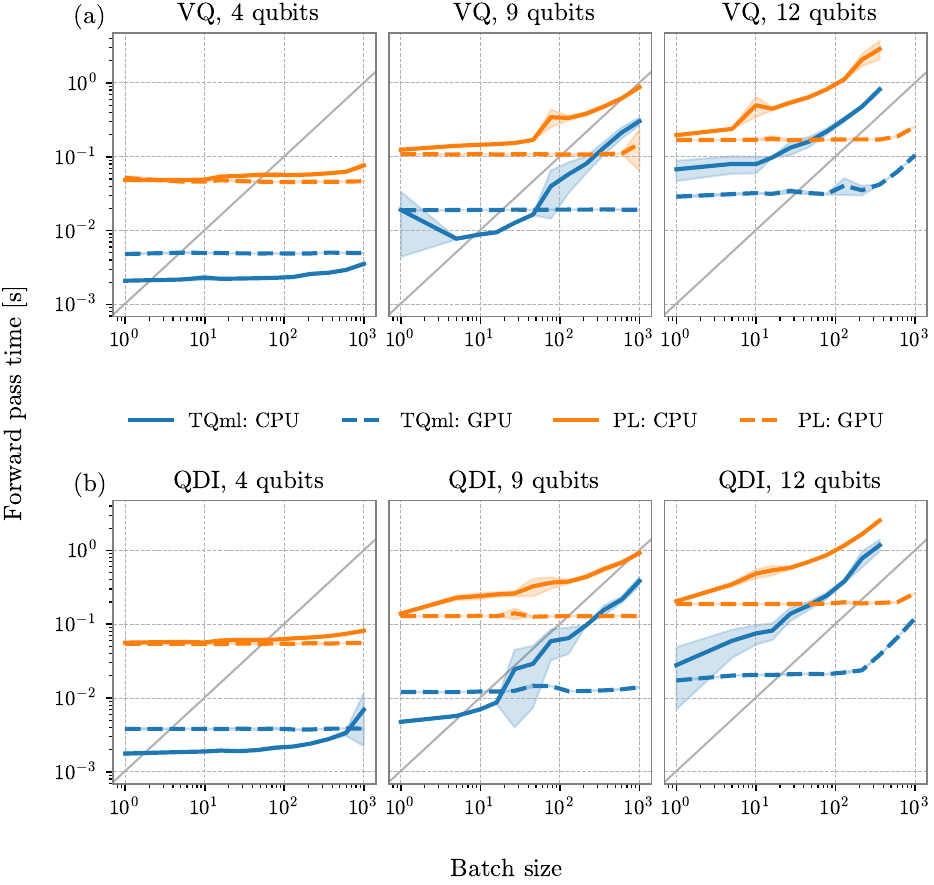}
\caption{Forward pass times for (a) Vanilla Quantum and (b) QDI layers, employing the circuit from Fig.~\ref{fig:Circuit_Rz_Ry_Cnot} for \(n=4\), 9, and 12 qubits, as a function of the batch size. Gray solid line indicates linear scaling. Comparisons are made between the PennyLane default.qubit simulator (PL) and our optimized TQml Simulator, with timings provided for one CPU thread and one GPU.}
    \label{fig:batching}
\end{figure}

The benchmark results in Fig.~\ref{fig:batching}—obtained using one CPU thread and one GPU—cover three system sizes, for each of which a different set of techniques is used for simulating the employed layers of gates. In all cases, our TQml Simulator shows a significant speedup compared to PennyLane. However, the speedup diminishes somewhat as the batch size approaches \(10^3\) (which is, though, a significant batch size in practice for the current stage of QML). Initially, the execution time does not increase with batch size, demonstrating effective hardware and software parallelization; however, at larger batch sizes, the scaling approaches linear scaling. As expected, the transition to linear scaling occurs slightly earlier for the QDI circuit than for the VQ circuit.

\begin{figure}[]
    \centering
    \includegraphics[width=5.9in]{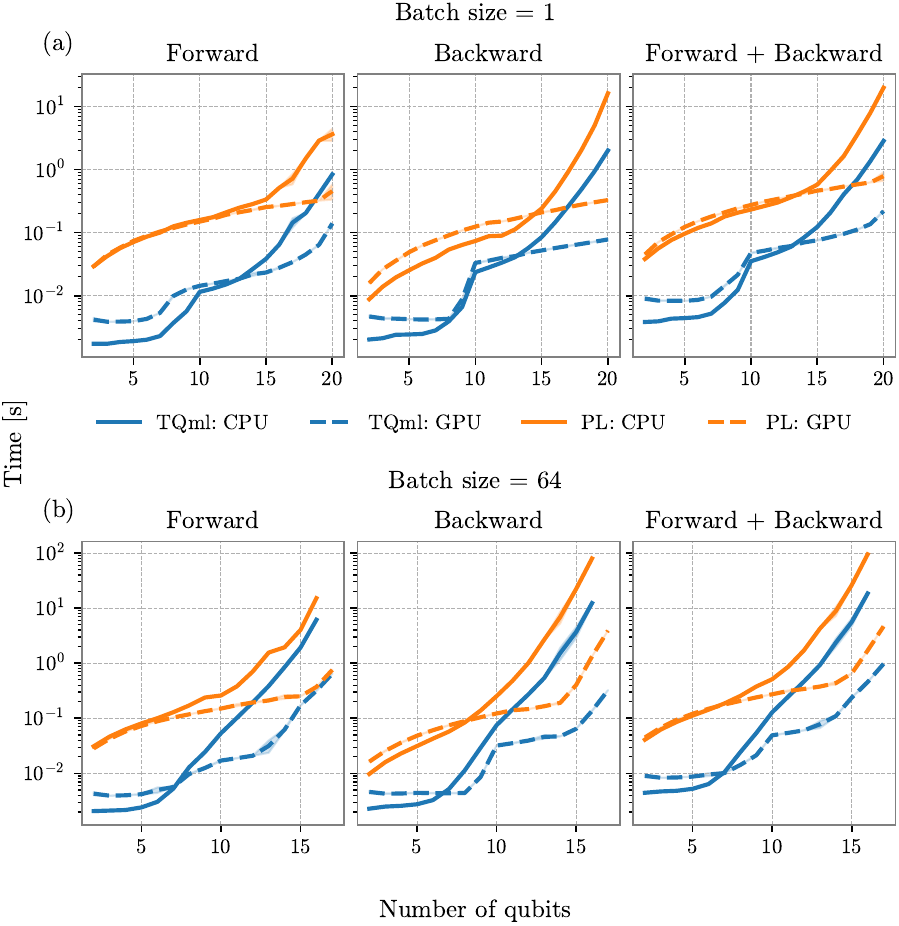}
\caption{Forward, backward, and total (summed) execution times for a QDI layer employing the circuit from Fig.~\ref{fig:Circuit_Rz_Ry_Cnot}, as a function of the number of qubits. Timings are provided for batch sizes of (a) 1 and (b) 64, using both the PennyLane default.qubit simulator (PL) and our optimized TQml Simulator, on one CPU thread and one GPU.}
    \label{fig:forward_and_backprop_scaling}
\end{figure}

Next, we benchmarked the forward, backward, and summed times for the QDI circuit in Fig.~\ref{fig:Circuit_Rz_Ry_Cnot} as a function of the number of qubits. This benchmark evaluates the performance of one routine iteration in the QML model training pipeline. For this test, we considered both one CPU thread and one GPU, and we examined batch sizes of 1 and 64. The results, shown in Fig.~\ref{fig:forward_and_backprop_scaling}, indicate that our TQml Simulator achieves an overall speedup of approximately one order of magnitude for systems with up to 10 qubits, with the speedup reducing to about half an order of magnitude for larger systems. Importantly, even though our optimality criterion was based solely on the forward pass time on a single CPU thread, the performance for backpropagation and GPU execution correlates well with this criterion, which is indicated by the absence of abrupt jumps at the points of the optimal technique changes.

\begin{figure}[]
    \centering
    \includegraphics[width=5.9in]{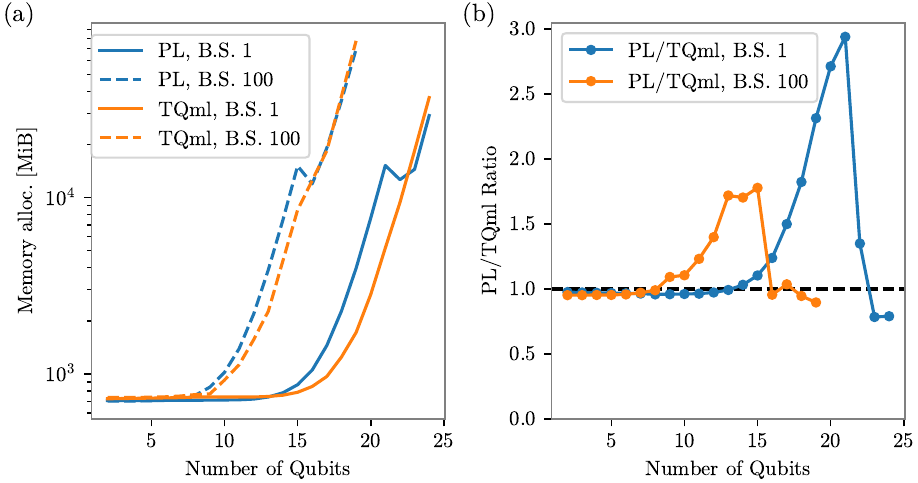}
\caption{(a) Memory allocation while simulating forward + backward passes of the QDI quantum layer with PennyLane default.qubit simulator (PL) and our optimized TQml Simulator on one CPU thread. The data is given for Batch Size (B.S.) 1 and 100. (b) Ratio of the memory allocations for PL and TQml simulators.}
    \label{fig:memory_allocation}
\end{figure}

We also benchmarked the memory allocation during the simulation of forward and backward passes for the QDI quantum layer. As shown in Fig.~\ref{fig:memory_allocation}, TQml Simulator exhibits a predictable, monotonically increasing memory footprint that scales exponentially with the number of qubits. In contrast, the memory allocation pattern of the default.qubit simulator is non-monotonic. As a result, our Optimized simulator outperforms default.qubit for circuits with up to a certain maximum number of qubits—depending on the batch size—but slightly underperforms when the system size exceeds this threshold. We leave the topic of memory optimization for future research.

\section{TQml Simulator with a JAX back-end}
\label{sec:tqml-jax}

    \begin{figure}[]
        \centering
        \includegraphics[width=5.9in]{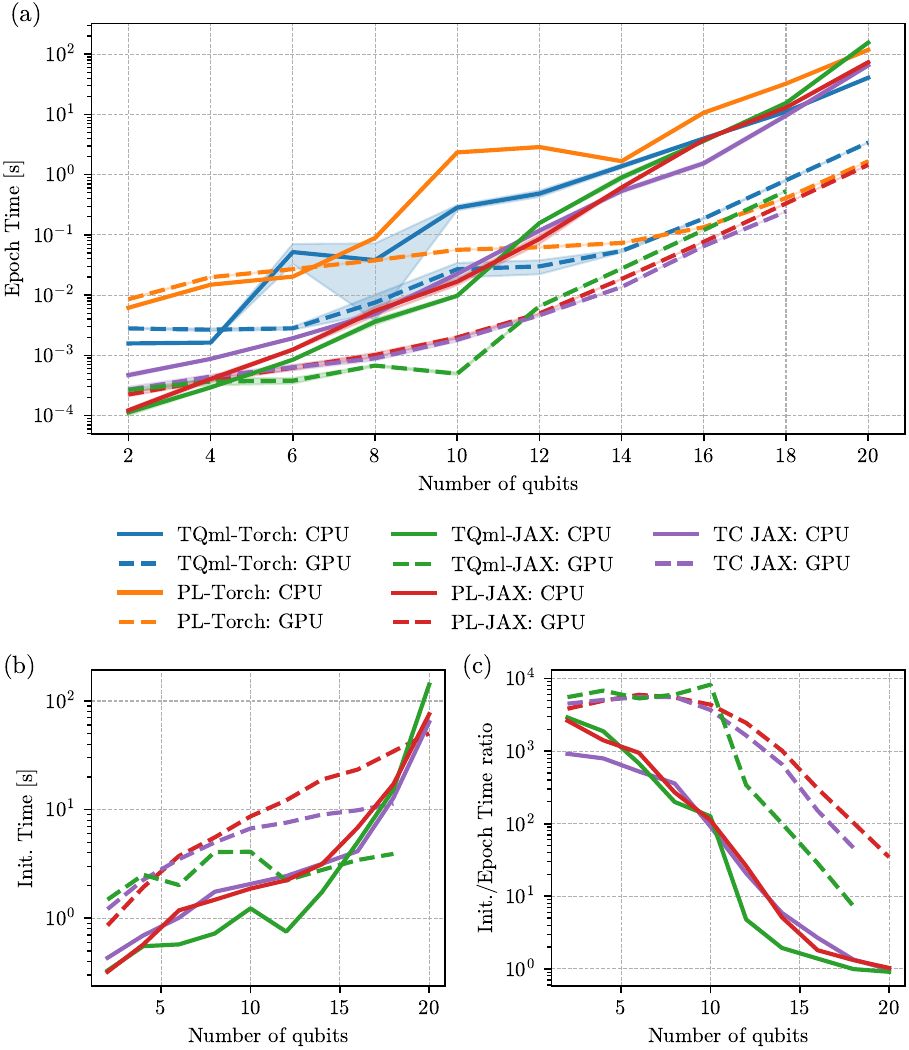}
\caption{(a) Forward{\,+}backward time per epoch for the Vanilla Quantum circuit (batch size~64) versus qubit number.  We compare PennyLane default.qubit (PL), TQml, and TensorCircuit (TC) on up to 48 CPU threads and on a single high-end GPU.  For PL and TQml both PyTorch and JAX back-ends are shown.  For the JAX runs the plotted time excludes the XLA compilation overhead.  
    (b) Wall-time for the \textit{first}, initialization, epoch with JIT compilation of the JAX simulator.  
    (c) Ratio between the first-epoch time and the steady-state epoch time for the JAX runs.}
        \label{fig:epoch_time_scaling_jax}
    \end{figure}
    
Replacing the PyTorch back–end of TQml with JAX is attractive because JAX provides a \emph{just-in-time} (JIT) decorator that hands the computation graph to XLA (Accelerated Linear Algebra), which in turn generates highly optimised machine code for the exact target hardware (CPU, GPU, or TPU).  The first time a function decorated with \texttt{@jax.jit} is used for a given set of argument shapes and dtypes, XLA compiles the graph; subsequent calls with the same signature re-use that compiled kernel and are therefore much faster.  The compilation step can be substantial, so it is essential to distinguish between the first epoch (which pays the compile cost) and later epochs (which do not).

We ported the TQml optimized for PyTorch and 1-thread CPU to JAX without changing the algorithmic choices of which techniques to use for simulating a particular layer for a given number of qubits.  For comparison, we benchmarked PennyLane \texttt{default.qubit},  TQml Simulators, and TensorCircuit~\cite{zhang2023tensorcircuit} (TC), which is based on tensor-network contractions, each in two variants when possible: PyTorch back-end and JAX back-end.  
Benchmarks were run on a 48-core CPU node (allowing each library to use all available threads) and on a single NVIDIA A100 GPU.  The circuit is a minor variant of Fig.~\ref{fig:Circuit_Rz_Ry_Cnot}; we focus on the Vanilla Quantum encoding and a batch size of~64.

Figure~\ref{fig:epoch_time_scaling_jax} (a) shows the steady-state epoch time (i.e.\ excluding JIT overhead for JAX).  Key observations are:

\begin{enumerate}
\item \textbf{Torch back-end.} 

    Tuned for a single-threaded CPU and Torch back-end, TQml-Torch consistently outperforms PennyLane-Torch by up to an order of magnitude.
    However, in this experiment's configuration, when being run on a GPU, PennyLane-Torch overtakes TQml-Torch at the highest qubit number.
    This crossover is consistent with our layer-selection having been calibrated for CPU rather than GPU; a GPU-specific re-optimisation of the choice of the simulation techniques should restore (and likely extend) TQml’s lead.
      
\item \textbf{JAX back-end.}  
      All three simulators benefit greatly from JIT compilation for the small number of qubits; after the first epoch their performances cluster together, indicating that XLA has efficiently optimized the computational graph down to fused BLAS kernels with similar performance. As the number of qubits increases, the performance gain, for CPU and GPU, diminishes, likely because the runtime becomes dominated by large matrix-matrix multiplications, which are already highly optimized in libraries like cuBLAS and MKL.
\end{enumerate}

Figure~\ref{fig:Circuit_Rz_Ry_Cnot} (b) and (c) quantify the price of JIT.  The first epoch is up to 4 orders of magnitude slower than later epochs, but this penalty shrinks with system size because the \emph{relative} weight of the compilation step decreases once the numerical linear-algebra dominates. Notably, TQml-JAX with GPU compiles in significantly less time than PennyLane-JAX or TensorCircuit-JAX with GPU.

While TQml-JAX is the fastest JAX-based simulator up to $\sim11$ qubits, its advantage diminishes for larger systems.  The reason is that we ported the set of per-layer techniques that is optimal for the PyTorch back-end; those choices are not necessarily optimal when XLA is free to fuse or reorder operations.  We again expect that re-optimising the techniques selection specifically for the JAX back-end is an obvious next step and should restore (or even extend) the performance gap at higher qubit counts.

\section{Conclusion}
\label{conclusion}

In this work, we investigated numerical simulation techniques for quantum machine learning (QML) circuits that consist of a set of methods to apply a layer of gates to a state-vector. We benchmarked these methods that leverage specific information about quantum gates—such as their layered presence in a circuit, locality, diagonality, real-valued nature, and effective permutation representations. We showed that the optimal simulation technique for a specific gate layer depends on the number of qubits. 

Building on these insights, we designed an optimized TQml Simulator that selects the most efficient simulation technique for each gate layer in a QML circuit. Implemented with a PyTorch back-end, our TQml Simulator was benchmarked against the PennyLane default.qubit simulator. The results demonstrate speedups by up to a factor of 10, depending on the circuit, the number of qubits, the batch size of the input data, and the hardware used. While the TQml Simulator demonstrated in our work selected methods that are optimal for a forward pass on a single CPU thread and PyTorch back-end, the approach is straightforwardly extended to other figures of merit that can take into account time for a backward pass, specific hardware used for simulation or memory allocation, or other back-ends such as JAX or CUDA Quantum~\cite{kim2023cuda}.

We note that while this work, among others, employs some basic tensor-network techniques—using Einsum to apply a layer of local gates—more advanced tensor-network approaches can be employed, such as those used in the TensorCircuit simulation library. Importantly, the techniques we consider for applying gates to a state vector can be straightforwardly extended to methods that contract gates with individual tensors within a tensor network. We leave this extension for future research.

\bibliographystyle{unsrt}   
\bibliography{main}         

\newpage

\appendix

\section{Appendix: Matrix Representations} \label{Matrix Representations}
Here, we provide matrix representations of quantum operations that include the gates considered in the main text. $ 2 \times 2 $ matrices and $ 4 \times 4$ matrices represent one-qubit and two-qubit gates, respectively. \\

\textbf{\textit{(Anti)diagonal:}}

\[
\begin{array}{|c|c|}
\hline
\textbf{Gate} & \textbf{Matrix Representation} \\
\hline
S & \begin{pmatrix} 1 & 0 \\ 0 & i \end{pmatrix} \\
\hline
Z & \begin{pmatrix} 1 & 0 \\ 0 & -1 \end{pmatrix} \\
\hline
Rz(\theta) & \begin{pmatrix} e^{-i\theta/2} & 0 \\ 0 & e^{i\theta/2} \end{pmatrix} \\
\hline
\text{CZ} & \begin{pmatrix} 1 & 0 & 0 & 0 \\ 0 & 1 & 0 & 0 \\ 0 & 0 & 1 & 0 \\ 0 & 0 & 0 & -1 \end{pmatrix} \\
\hline
Rzz(\theta) & \begin{pmatrix} e^{-i\theta/2} & 0 & 0 & 0 \\ 0 & e^{i\theta/2} & 0 & 0 \\ 0 & 0 & e^{i\theta/2} & 0 \\ 0 & 0 & 0 & e^{-i\theta/2} \end{pmatrix} \\
\hline
\text{GPI}(\phi) & \begin{pmatrix} 0 & e^{-2\pi i \phi} \\ e^{2\pi i \phi} & 0 \end{pmatrix} \\
\hline
\end{array}
\]

\textbf{\textit{Permutation:}}

\[
\begin{array}{|c|c|}
\hline
\textbf{Gate} & \textbf{Matrix Representation} \\
\hline
X & \begin{pmatrix} 0 & 1 \\ 1 & 0 \end{pmatrix} \\
\hline
\text{SWAP} & \begin{pmatrix} 1 & 0 & 0 & 0 \\ 0 & 0 & 1 & 0 \\ 0 & 1 & 0 & 0 \\ 0 & 0 & 0 & 1 \end{pmatrix} \\
\hline
\text{CNOT} & \begin{pmatrix} 1 & 0 & 0 & 0 \\ 0 & 1 & 0 & 0 \\ 0 & 0 & 0 & 1 \\ 0 & 0 & 1 & 0 \end{pmatrix} \\
\hline
\end{array}
\]

\textbf{\textit{Fixed gates:}}

\[
\begin{array}{|c|c|}
\hline
\textbf{Gate} & \textbf{Matrix Representation} \\
\hline
H & \displaystyle \frac{1}{\sqrt{2}}
\begin{pmatrix}
1 & 1 \\
1 & -1
\end{pmatrix} \\
\hline
S_x & \displaystyle \frac{1}{2}
\begin{pmatrix}
1 + i & 1 - i \\
1 - i & 1 + i
\end{pmatrix} \\
\hline
\text{ECR} & \displaystyle \frac{1}{\sqrt{2}}
\begin{pmatrix}
0 & 1 & 0 & i \\
1 & 0 & -i & 0 \\
0 & 0 & i & 0 \\
-i & 0 & 0 & 1
\end{pmatrix} \\
\hline
\end{array}
\]

\textbf{\textit{Single qubit rotation:}}

\[
\begin{array}{|c|c|}
\hline
\textbf{Gate} & \textbf{Matrix Representation} \\
\hline
R_x(\theta) & \displaystyle \begin{pmatrix} \cos\frac{\theta}{2} & -i\sin\frac{\theta}{2} \\ -i\sin\frac{\theta}{2} & \cos\frac{\theta}{2} \end{pmatrix} \\
\hline
R_y(\theta) & \displaystyle \begin{pmatrix} \cos\frac{\theta}{2} & -\sin\frac{\theta}{2} \\ \sin\frac{\theta}{2} & \cos\frac{\theta}{2} \end{pmatrix} \\
\hline
\text{Rot}(\theta,\phi,\lambda) & \displaystyle \begin{pmatrix} \cos\frac{\theta}{2} & -e^{i\lambda}\sin\frac{\theta}{2} \\ e^{i\phi}\sin\frac{\theta}{2} & e^{i(\phi+\lambda)}\cos\frac{\theta}{2} \end{pmatrix} \\
\hline
\end{array}
\]

\textbf{\textit{Other gates:}}

\begin{table}[ht]
    \centering
    \resizebox{\textwidth}{!}{%
    \begin{tabular}{|c|c|}
        \hline
        \textbf{Gate} & \textbf{Matrix Representation} \\
        \hline
        \text{GPI2}($\phi$) & 
        $\displaystyle \frac{1}{\sqrt{2}}
        \begin{pmatrix}
            1 & -\,i\,e^{-2\pi i \phi} \\
            -\,i\,e^{2\pi i \phi} & 1
        \end{pmatrix}$ \\
        \hline
        \text{MS}($\phi_0, \phi_1, \theta$) &
        $\begin{pmatrix}
            \cos(\pi \theta) & 0 & 0 & -\,i\,e^{-2\pi i (\phi_0 + \phi_1)} \,\sin(\pi \theta) \\
            0 & \cos(\pi \theta) & -\,i\,e^{-2\pi i (\phi_1 - \phi_0)} \,\sin(\pi \theta) & 0 \\
            0 & -\,i\,e^{-2\pi i (\phi_0 - \phi_1)} \,\sin(\pi \theta) & \cos(\pi \theta) & 0 \\
            -\,i\,e^{-2\pi i (\phi_0 + \phi_1)} \,\sin(\pi \theta) & 0 & 0 & \cos(\pi \theta)
        \end{pmatrix}$ \\
        \hline
    \end{tabular}
    }
\end{table}

\section{Hardware-specific Simulators: IBM and IonQ}

In this section, we present numerical comparisons of the performance of our TQml Simulator—designed to minimize the runtime for each individual gate layer—for circuits built with native gates specific to IBM superconducting and IonQ ion-trap quantum computers. These results are compared with those obtained using the PennyLane default.qubit simulator.

\label{Hardware-specific simulators}

    \begin{figure}[]
        \centering
        \includegraphics[width=0.8\linewidth]{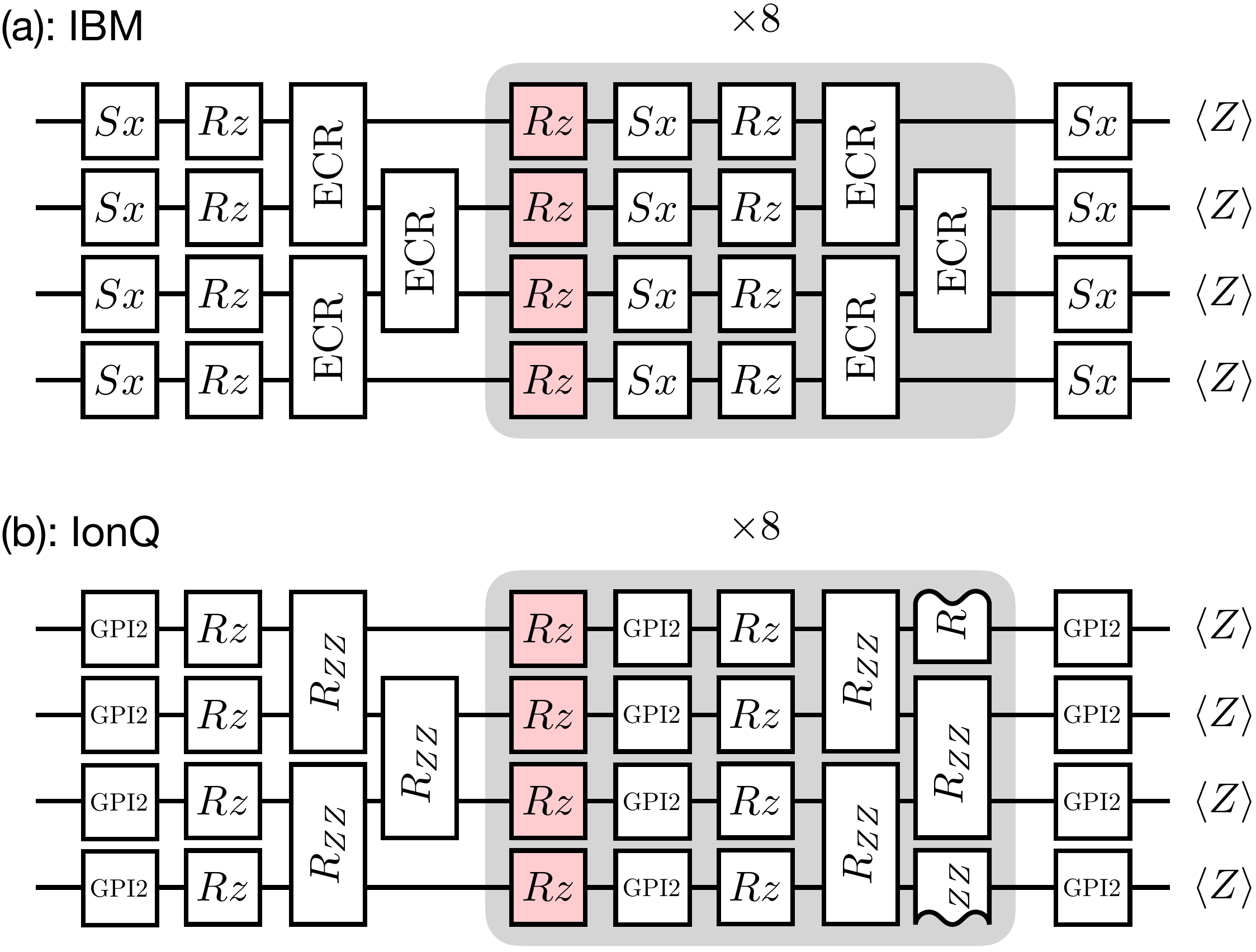}
        
    \caption{4-qubit example of a circuit built with (a) IBM's Eagle-type superconducting processor native gates and (b) IonQ's Forte-type ion-trap processor native gates used for benchmarks reported in Figs.~\ref{fig:forward_and_backprop_scaling_IBM} and \ref{fig:forward_and_backprop_scaling_IonQ}. The shaded area is repeated 8 times. The layer of \(R_z\) gates highlighted in red is used for encoding. To yield the output and its derivatives, single-qubit observables \(\langle Z \rangle\) are measured and summed.}
        \label{fig:IBM_IonQ_circuits}
    \end{figure}

    \begin{figure}[]
        \centering
        \includegraphics[width=5.9in]{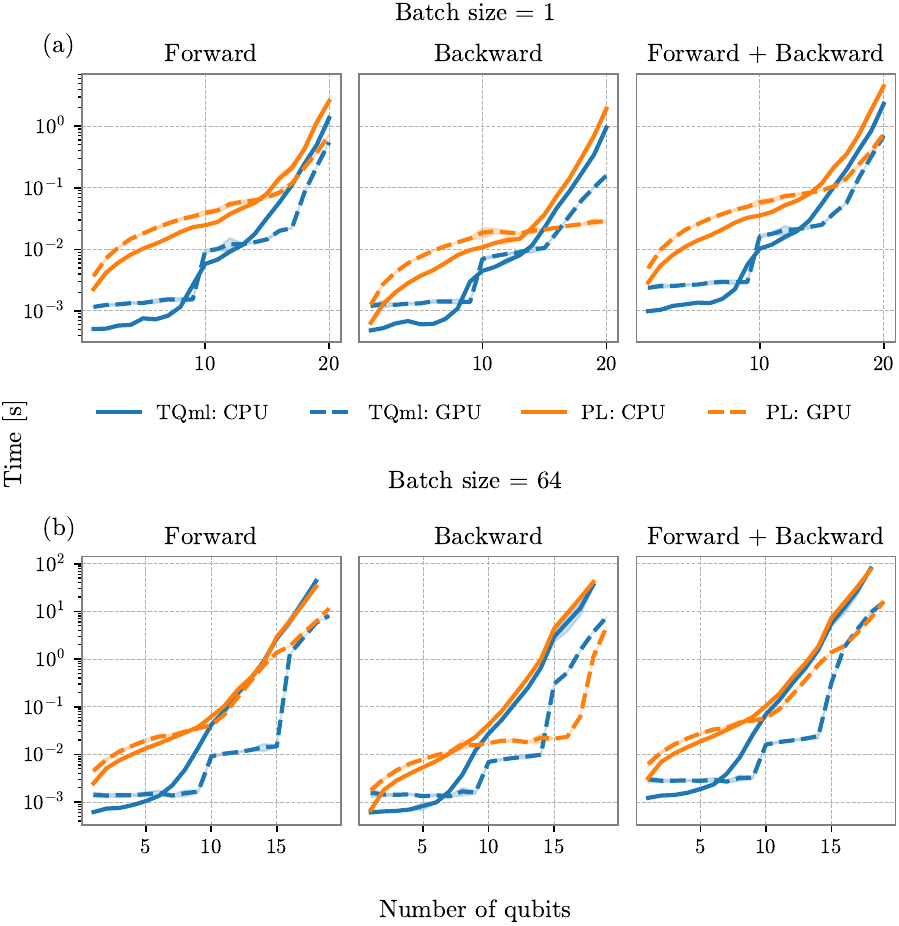}
        \caption{Forward, backward, and total (summed) execution times for a QDI layer employing the circuit from Fig.~\ref{fig:IBM_IonQ_circuits}(a) build with IBM's Eagle superconducting processors native gates, as a function of the number of qubits. Timings are provided for batch sizes of (a) 1 and (b) 64, using both the PennyLane default.qubit simulator (PL) and our optimized TQml Simulator, on one CPU thread and one GPU.}
        \label{fig:forward_and_backprop_scaling_IBM}
    \end{figure}
    
Native gates on IBM's Eagle-type superconducting processors include ECR, X, Sx, and \(R_z\). Using these gates, we constructed a QDI circuit layer~\cite{sagingalieva2023hybrid, sagingalieva2025photovoltaic} (see also the main text) as shown in Fig.~\ref{fig:IBM_IonQ_circuits}(a). The benchmark results presented in Fig.~\ref{fig:forward_and_backprop_scaling_IBM} demonstrate that the TQml Simulator achieves a speedup of up to an order of magnitude for circuits with up to approximately 15 qubits, depending on the batch size and the hardware used for simulation. However, as the number of qubits increases, the performance advantage of our simulator converges towards that of the default.qubit simulator.    

    \begin{figure}[]
        \centering
        \includegraphics[width=5.9in]{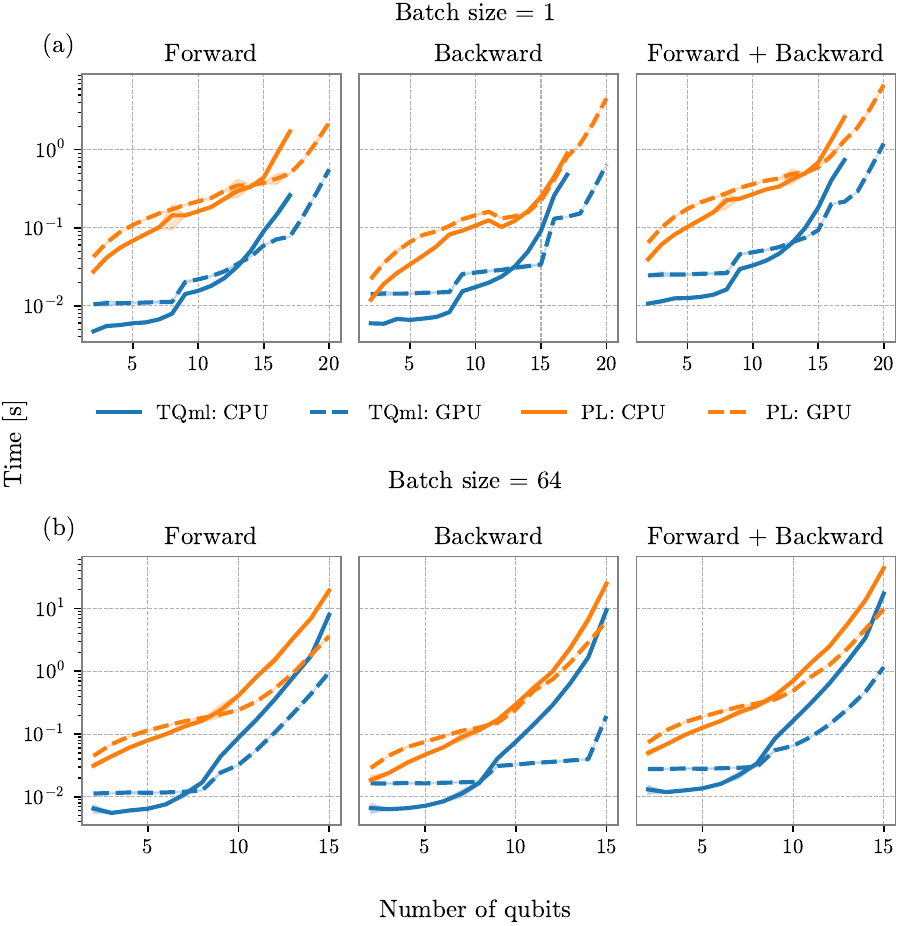}
        \caption{Forward, backward, and total (summed) execution times for a QDI layer employing the circuit from Fig.~\ref{fig:IBM_IonQ_circuits}(a) build with IonQ's Forte ion-trap processors native gates, as a function of the number of qubits. Timings are provided for batch sizes of (a) 1 and (b) 64, using both the PennyLane default.qubit simulator (PL) and our optimized TQml Simulator, on one CPU thread and one GPU.}
        \label{fig:forward_and_backprop_scaling_IonQ}
    \end{figure}
    
Native gates on IonQ's Forte-type ion-trap processors include GPI, GPI2, Rzz, and virtual \(R_z\). For these processors, we constructed a QDI circuit layer as illustrated in Fig.~\ref{fig:IBM_IonQ_circuits}(b). The results in Fig.~\ref{fig:forward_and_backprop_scaling_IonQ} show a more consistent speedup of the TQml Simulator—up to an order of magnitude—compared to the default.qubit simulator. The maximum number of qubits that can be simulated is constrained by the available memory on the test machine. For a batch size of 1, the default.qubit simulator reached 17 qubits, whereas our TQml Simulator was able to simulate up to 20 qubits on the same machine.

\end{document}